\documentclass[10pt,a4paper]{article}

\usepackage{amssymb}
\usepackage{amsmath}
\usepackage{bm}
\usepackage{graphicx}
\usepackage{subfigure}
\usepackage{hyperref}
\usepackage{color,soul}

\usepackage{booktabs}
\usepackage{array}
\usepackage{enumitem}
\usepackage{cite}
\usepackage[affil-it,auth-lg]{authblk}
\usepackage{eso-pic}

\begin{document}
	
	\title{Gravitational potential from maximum entropy principle}
	\author[1]{Zacharias Roupas}
	\affil[1]{Centre for Theoretical Physics, The British University in Egypt, Sherouk City 11837, Cairo, Egypt} 
	
	\date{\vspace{-5ex}}
	
	\maketitle

\begin{abstract}
{It is shown here} in the framework of standard General Relativity that the gravitational potential in static spacetime, equivalently the redshift factor, inside any kind of matter, can be derived from maximum entropy principle. It is used only the Hamiltonian constraint, without further invoking Einstein's equations or any new principle. The Newtonian potential arises from the same procedure.  
\end{abstract}

\section{Introduction}

A deep connection between gravity and thermodynamics was for the first time hinted by the discovery of the four laws of black hole mechanics \cite{PhysRevD.7.2333,Bardeen1973} and Hawking radiation \cite{Hawking1975,PhysRevD.14.870}. The inclusion of quantum effects was required in order for such connection to be revealed. Later, several attempts were made to realize gravity as an effective, emergent theory and derive it from thermodynamics \cite{PhysRevLett.75.1260,Padmanabhan_2010,Verlinde2011}. In these formulations the key ingredient was the introduction and use of a new principle,  that was in all cases some form of holographic principle \cite{Susskind_1995,Bousso_2002}, which connected the quantum and relativistic realms and allowed for gravity to emerge as the favored maximum entropy configuration of underlying degrees of freedom.

Nevertheless, especially because of the effective nature of the energy-momentum tensor, thermodynamics and gravitational effects intervene even at the level of standard General Relativity without invoking any additional principle.
The maximum entropy principle in standard General Relativity has been used by several authors \cite{Cocke_1965,sorkin,bvgrg,Gao:2011hh,Fang_2014PhRvD..90d4013F,2014CQGra..31e5003S} in the derivation of relativistic hydrostatic equilibrium, namely Tolman-Oppenheimer-Volkoff (TOV) equation. In these studies, the Tolman law, continuity equation or a certain equation of state was pre-assumed. On the other hand, the Tolman law was derived together (and not preassumed) with TOV equation from maximum entropy principle in References \cite{Roupas_2013CQGra..30k5018R,Roupas_2015CQGra..32k9501R}. 

Deepening our understanding of the role of thermodynamics within standard general relativity will allow to achieve progress in the general research program of emergent gravity by dissociating the parts of the theory that require the introduction of additional principles from the ones which do not. Here, {we} demonstrate a calculation which leads to a rather unexpected result. {We} will derive the gravitational potential, namely the redshift factor, inside matter without pre-assuming Tolman law, continuity equation or TOV equation, and for any equation of state of matter. 

In standard General Relativity,  for a static spacetime in the presence of matter the redshift factor may be expressed with respect to the pressure and mass distribution by solving the Einstein's equations \cite{Weinberg_1972gcpa.book} (see Appendix).
 
Tolman \cite{Tolman:1930}, showed back in 1930 that, in the presence of matter, thermodynamic equilibrium requires temperature $T(\bm{r})$, as measured by a local observer, to be inhomogeneously distributed, generating a temperature gradient. He showed that in General Relativity, the thermodynamic quantity that is constant in equilibrium is 
$
	T(\bm{r})\sqrt{g_{tt}(\bm{r})} = const.
$
This looks a lot like a gravitational redshift formula. Nevertheless, it should be emphasized that it is deduced from a maximum entropy principle \cite{Tolman-Ehrenfest:1930} and as such its physical origin is different. 

{One may wonder} if it is possible to deduce $g_{tt}$ itself from maximum entropy principle without invoking Einstein's equations or continuity equation. {We} find that this is at least partially possible. In addition to maximum entropy principle, the Hamiltonian constraint should be used. This accounts for a-priori knowledge of a certain expression for the proper spatial volume, i.e. knowing the spatial 3-metric. 

\section{Maximum Entropy Principle}

For a static spacetime in {Schwarzschild} coordinates the metric may be written as \cite{Weinberg_1972gcpa.book}
\begin{equation}\label{eq:metric}
	ds^2  = -g_{tt}(r)c^2 dt^2 + g_{rr}(r)dr^2 + r^2d\Omega,
\end{equation}
where $d\Omega = d\theta^2 + sin^2\theta d\phi^2$ is the solid angle. We assume we know the spatial component $g_{rr}$ from the Hamiltonian constraint and wish to infer the time component from maximum entropy principle. 

The Hamiltonian constraint is
\begin{equation}
	^{(3)}R = \frac{16\pi G}{c^2} \rho
\end{equation}
where 
$
	^{(3)}R = \frac{2}{r^2}\frac{d}{dr}\left(r(1-g_{rr}^{-1})\right)
$
is the spatial intrinsic Ricci scalar. 
It gives 
\begin{equation}\label{eq:g_rr}
	g_{rr} = \left(1 - \frac{2GM(r)}{c^2} \right)^{-1}
\end{equation}
with
\begin{equation}
	\frac{d M(r)}{dr} = 4\pi r^2 \rho (r). 
\end{equation}
In this respect, $\rho(r)$ is the total mass-energy density of the system at $r$ including gravitational degrees of freedom \cite{Weinberg_1972gcpa.book}. $M(r)$ is the  total mass-energy enclosed within $r$.
In the followings, let us use the function
\begin{equation}
	f(r,M(r)) = \sqrt{g_{rr}}.
\end{equation}
This will allow possible generalizations and help us gain further insight into the calculation.

Suppose we observe gravitational effects which we measure by some energy (volume) density of gravitational degrees of freedom, that we denote $\varepsilon_G$. Let us further denote $n_q(r) = dN_q(r)/dV$ the particles' number density, where $q$ counts particles' species, each of mass $m_q$.
Assume we wish to model the effects of gravity in the spherically symmetric static case in the presence of matter with some unknown function $f(r,M(r)) \geq 1$, which does not rearrange matter at infinity $f(r\rightarrow\infty)=1$, as follows
\begin{equation}\label{eq:e_G}
	\varepsilon_G(r) = \rho(r)c^2 - f(r,M) \sum_q m_q c^2 n_q(r).
\end{equation}
This equation implies that if after the subtraction of rest mass energy from total mass-energy you are left with non-zero, negative energy, then gravity is present. Its presence is accounted for by invoking the function $f$. 
Taking correctly into account the redistribution of matter due to gravity, the total number of particles of the $q$th specie is 
\begin{equation}
	N_q = \int_0^R dr\,4\pi r^2 f(r,M) n_q(r) ,
\end{equation}
where $R$ is the radius up to which matter $n_q$ is present.
Likewise the total entropy of the system within this radius is 
\begin{equation}
	S = \int_0^R dr\, 4\pi r^2 f(r,M) s(r),
\end{equation}
where $s = s(r)$ is the entropy density. Acknowledging the relation between $f$ and $g_{rr}$ we understand that the integrals are actually performed over the proper 3-volume $dV_\text{prop} = \sqrt{g_{rr}}dV$. The total mass-energy of the system up to $R$ is simply
\begin{equation}
	\mathcal{M}_\text{tot} = \int_0^R dr\,4\pi r^2 \rho(r). 
\end{equation}

We also assume that thermodynamics holds. The first law for spatial densities is expressed locally \cite{1973grav.book.....M} as 
\begin{equation}\label{eq:1st_law}
	T(r)ds(r) = d\rho(r) c^2 - \sum_q\mu_q (r) dn_q(r).
\end{equation}
This equation implies that in the most general case
\begin{equation}\label{eq:s_dep}
	s = s(\rho,n_q).
\end{equation}
{We} use units where $k=1$ so that temperature $T$ is measured in energy units and entropy is dimensionless. 
Beware that we allow for local temperature $T$ and chemical potential $\mu_q$ to depend on the position in equilibrium. The precise dependence is exactly what we wish to derive. There is a second equation involving thermodynamic densities, called Euler equation, or sometimes integrated Gibbs-Duhem relation,
\begin{equation}\label{eq:Euler}
	T(r)s(r) = P(r) + \rho(r) c^2 - \sum_q\mu_q(r) n_q(r),
\end{equation}
where $P$ is the pressure. We do not assume a specific equation of state.

We invoke the 2nd law of thermodynamics for constant total mass-energy $\mathcal{M}_\text{tot}$ and number of particles $N_q$, introducing the Lagrange multipliers
\begin{equation}
	\beta = const.,\;
	a_q = const.
\end{equation}
We have
\begin{equation}\label{eq:dS_1}
	\delta 	S - \beta \delta \mathcal{M}_\text{tot} c^2 + \sum_q a_q \delta N_q = 0,
\end{equation}
where the entropy is assumed to be a functional 
\begin{equation}
	S = S[\rho,n_q]
\end{equation}
through Eq. (\ref{eq:s_dep}). 

We get from Eq. (\ref{eq:dS_1}), by use of Eq.(\ref{eq:1st_law}), that
\begin{equation}
\int_0^R dr\, 4\pi r^2 \left\lbrace 
	\left( s + \sum_q a_q n_q\right) \delta f 
	+ f \sum_q\left(a_q - \frac{\mu_q}{T}\right)\delta n_q 
	+ c^2 \left( \frac{f}{T} - \beta\right)\delta \rho 
	\right\rbrace
	= 0.
\end{equation}
Since $\delta n_q$ are varied independently we deduce
\begin{equation}\label{eq:mu_T}
	\frac{\mu_q(r)}{T(r)} = a_q = const.
\end{equation}
That is, the chemical potential at thermal equilibrium attains the same profile as the temperature.
We get substituting (\ref{eq:mu_T}) and using also Euler Eq. (\ref{eq:Euler})
\begin{equation}
		\int_0^R dr\, 4\pi r^2 \left\lbrace
	\frac{ P + \rho c^2}{T} \delta f
	+ c^2 \left( \frac{f}{T} - \beta\right)\delta \rho 
	\right\rbrace = 0.
\end{equation}
At this point note that $\delta f(r)$ is not independently varied, since it depends on $\delta \rho(r)$ through $\delta M(r)$
\begin{equation}\label{eq:df}
	\delta f (r) = \frac{\partial f}{\partial M}\int_0^r dx\, 4\pi x^2 \delta \rho(x).
\end{equation} 

Substituting (\ref{eq:df}) we get 
\begin{equation}
\int_0^R dr\, \left\lbrace
		4\pi r^2 
	\frac{ P(r) + \rho(r) c^2}{T(r)} \frac{\partial f(r)}{\partial M(r)} \int_0^r dx\, 4\pi x^2 \delta \rho(x)
	\right\rbrace+ 
	\int_0^R dr\, 4\pi r^2 c^2 \left( \frac{f}{T} - \beta\right)\delta \rho
	= 0.
\end{equation}
The first term is a double integral whose integration order we wish to change. We have for any function $h = h(x,r)$ that 
\begin{equation}
\int_0^R dr \left(\int_0^r dx\, h(x,r)\right) = 
\int_0^R dx \left(\int_{x}^R dr\, h(x,r)\right)
=	\int_0^R dr \left(\int_{r}^R dx\, h(r,x)\right).
\end{equation}
since
$\{0 \leq r \leq R ,0 \leq x \leq r\}
\Rightarrow
\{0 \leq x \leq R,x \leq r \leq R\}
$.
Changing the integration order we get the value of the other Lagrange multiplier
\begin{equation}\label{eq:beta}
	\beta = \frac{f}{T} + \int_r^R dx\, 4\pi x^2 \frac{ \frac{P}{c^2} + \rho}{T} \frac{\partial f}{\partial M} .
\end{equation}
Note that
\begin{equation}
	T(R) = f(R) \beta^{-1}\quad \text{and}\quad
	T(r>R) = f(r)\beta^{-1},
\end{equation}
and 
\begin{equation}
	f(r\rightarrow \infty) =  1 \Rightarrow T_\infty = \beta^{-1}.
\end{equation}
This is the physical interpretation of the Lagrange multiplier $\beta$. It is the inverse temperature of the mass $\mathcal{M}_\text{tot}$, enclosed in $R$, measured by an observer at infinity.
Taking the derivative of expression (\ref{eq:beta}) and considering that
\begin{equation}
	\frac{df}{dr} = \frac{\partial f}{\partial M} \frac{dM}{dr} + \frac{\partial f}{\partial r}
	= \frac{\partial f}{\partial M} 4\pi r^2\rho + \frac{\partial f}{\partial r},
\end{equation}
we get
\begin{equation}\label{eq:dlnT}
	\frac{d\ln T}{dr} = -\left( 4\pi r^2 \frac{P}{c^2}\frac{\partial f}{\partial M} - \frac{\partial f}{\partial r}\right) f^{-1}.
\end{equation}
Integrating from $r$ to infinity  we get
\begin{align}
\nonumber	T(r) \exp &\left\lbrace -\int_r^\infty dx\, \left( 4\pi x^2 \frac{P}{c^2}\frac{\partial f}{\partial M} - \frac{\partial f}{\partial x} \right) f^{-1} \right\rbrace = \\
\label{eq:T_r}	&= \beta^{-1} = const.
\end{align}
If now we identify
\begin{equation}\label{eq:g_tt_gen}
g_{tt}  = \exp\left\lbrace
-2\int_r^\infty dx\,\left( 4\pi x^2 \frac{P}{c^2}\frac{\partial f}{\partial M} - \frac{\partial f}{\partial x} \right)  f^{-1} \right\rbrace
\end{equation}
such that it satisfies Eq. (\ref{eq:metric}), and identify also $g_{rr} = f^2$, we recover spacetime geometry as well as Tolman law
\begin{equation}
T(r)\sqrt{g_{tt}} = \beta^{-1} = const.
\end{equation}

If we further substitute in $f$ the value of $g_{rr}$, Eq. (\ref{eq:g_rr}), given by the Hamiltonian constraint we get
\begin{equation}\label{eq:g_tt}
g_{tt} = \exp\left\lbrace-\frac{2}{c^2}\int_r^\infty dx\, 
\left(\frac{GM}{x^2} + \frac{4\pi G}{c^2} P\, x\right)\left( 1 - \frac{2G M(x) }{x c^2} \right)^{-1} \right\rbrace.
\end{equation}
This is indeed the correct redshift factor as in Eq. (\ref{eq:nu}). Using Eqs. (\ref{eq:1st_law}), (\ref{eq:Euler}), (\ref{eq:mu_T}) we get $P' = (P+\rho c^2) T'/T$. Substituting $T'$ we can solve (the resulting TOV equation) for $P(r)$, $M(r)$ given an equation of state. Then, we substitute in (\ref{eq:g_tt}) and get $g_{tt}$.

Apparently, what we achieved is to calculate $g_{tt}(r)$ with sole input the spatial component $g_{rr}(r)$ and thermodynamics!

Now, if we define
\begin{equation}\label{eq:phi}
	\phi(r) = - c^2 \ln\frac{T(r)}{T_\infty},
\end{equation}
then Eq. (\ref{eq:dlnT}) gives
\begin{equation}
	\frac{d\phi(r)}{dr} = 	\left\lbrace
	\begin{array}{ll}
\left(\frac{GM}{r^2} + \frac{4\pi G}{c^2} P\, r\right)\left( 1 - \frac{2G M(r) }{r c^2} \right)^{-1},& r\leq R \\[2ex]
	G\frac{\mathcal{M}_\text{tot}}{r^2} \left( 1 - \frac{2G \mathcal{M}_\text{tot} }{r c^2} \right)^{-1},& r\geq R .
	\end{array}
	\right.
\end{equation}
Thus, the quantity $\phi(r)$ defined in (\ref{eq:phi}) may be regarded as the relativistic generalization of a Newtonian gravitational potential. Indeed we get for non-relativistic matter and weak field
\begin{equation}
	\frac{d\phi(r)}{dr} \simeq \frac{GM}{r^2},\quad \mbox{for}\; P\ll c^2\;\mbox{and}\; GM\ll rc^2.
\end{equation}
We also deduce for the relativistic generalization of Newtonian potential \cite{Wald:book} from Eq. (\ref{eq:g_tt}) that $g_{tt} = e^{2\phi/c^2}$.

The Newtonian potential is tightly connected with temperature as implied by Eq. (\ref{eq:phi}). Under the  perspective of this work, it is an $\mathcal{O}(c^{-2})$ effect \footnote{This is also understood already from the Einstein's equations in the weak field limit 
$
	G_{00} = \frac{8\pi G}{c^4}T_{00}\Rightarrow
	\nabla^2 g_{00} = \frac{8\pi G}{c^2}\rho . 
$
The metric has to be approximated to $\mathcal{O}(c^{-2})$ in order for the Newtonian potential to make its presence evident $g_{00} = 1 + 2\phi c^{-2} + \mathcal{O}(c^{-4}).
$ Then, $c^{-2}$ drops eventually from the Einstein's equation which become the Poisson equation.}.
Although the Tolman effect is negligible in most astrophysical settings, Newtonian gravity emerges together with it. 

\section{Conclusions}

It is shown here how to deduce the gravitational potential $\phi(r)$, equivalently the redshift factor $g_{tt}(r)=e^{2\phi(r)/c^2}$, inside matter, given the spatial metric, using maximum entropy principle and the Hamiltonian constraint in standard General Relativity without invoking any additional principle. Since the redshift is related to the Newtonian potential, the later is deduced by the same procedure, as in equation  (\ref{eq:phi}). 

This work indicates the presence of an implicit thermodynamic sector in the theory of General Relativity, which {may} require further investigation. It designates a thermodynamic property of General Relativity that any other alternative theory of gravity should also respect. Therefore it provides a thermodynamic consistency check and can serve as a selection criterion for good modified theories of gravity.

\appendix   

\section{Redshift factor from Einstein's equations}\label{app:Ein_lapse}

The metric of a spherically symmetric and static spacetime may be written in Schwarzschild coordinates as \cite{Weinberg_1972gcpa.book}
\begin{equation}
	ds^2 = - e^{\nu(r)}c^2dt^2 + e^{\lambda (r)} dr^2 + r^2d\Omega^2.
\end{equation}
The Einstein's equations give 
\begin{eqnarray}
\label{eq:tensorcom1}	\frac{8\pi G}{c^4}T^0_0 &=& e^{-\lambda}\left( \frac{\lambda '}{r} - \frac{1}{r^2}\right) + \frac{1}{r^2}
\\
\label{eq:tensorcom2}	\frac{8\pi G}{c^4}T^1_1 &=& -e^{-\lambda}\left( \frac{\nu '}{r} + \frac{1}{r^2}\right) + \frac{1}{r^2} 
 \\
\label{eq:tensorcom3}	
\frac{8\pi G}{c^4}T^2_2 &=& -e^{-\lambda}\left( \frac{\nu ''}{2} - \frac{\lambda'\nu'}{4} + \frac{{\nu '}^2}{4} + \frac{\nu'-\lambda'}{2r}\right) 
\end{eqnarray}
where prime denotes differentiation with respect to $r$. The energy-momentum tensor of a perfect fluid at equilibrium is
\begin{equation}
	T^{\mu}_\nu = \text{diag}(-\rho(r) c^2,P(r),P(r),P(r)),
\end{equation}
where $\rho$ is the mass-energy density and $P$ the pressure. Therefore equations (\ref{eq:tensorcom1})-(\ref{eq:tensorcom3}) become
\begin{align}
\label{eq:tolequi1}	\frac{dP}{dr} &= - \frac{\nu'}{2}(P + \rho c^2) 
\\
\label{eq:tolequi2}	\frac{8\pi G}{c^2}\rho &= e^{-\lambda}\left( \frac{\lambda '}{r} - \frac{1}{r^2}\right) + \frac{1}{r^2} 
\\
\label{eq:tolequi3}	\frac{8\pi G}{c^4} P  &= e^{-\lambda}\left( \frac{\nu'}{r} + \frac{1}{r^2}\right) - \frac{1}{r^2}. 
\end{align}
Equation (\ref{eq:tolequi1}) is derived by equating (\ref{eq:tensorcom2}) and (\ref{eq:tensorcom3}). Applying the transformation
\begin{equation}
	e^{-\lambda} = 1 - \frac{2GM(r)}{r}
\end{equation}
to equation (\ref{eq:tolequi3}) and solving with respect to $\nu'$ we calculate the redshift factor
\begin{equation}\label{eq:nu}
	\nu = -\frac{2}{c^2}\int_r^\infty dr \left(\frac{G M(r)}{r^2} + 4\pi G \frac{P}{c^2}r \right)\left(1 - \frac{2G M(r)}{rc^2}\right)^{-1},
\end{equation}
with
\begin{equation}
	g_{tt} = e^\nu.
\end{equation}
The remaining equation (\ref{eq:tolequi2}) gives
\begin{equation}\label{eq:mprime}
	M' = 4\pi\rho r^2.
\end{equation}

Now, substituting $\nu'$ into equation (\ref{eq:tolequi1}), we get TOV equation:
\begin{equation}\label{eq:TOV1}
	P' = - \left(\frac{P}{c^2} + \rho\right) \left( \frac{G M(r)}{r^2} + 4\pi G \frac{P}{c^2}r \right)\left(
	 1 - \frac{2G M(r)}{rc^2} \right)^{-1}	
\end{equation}
that is the equation of relativistic hydrostatic equilibrium. Given an equation of state, one can calculate $P$, $M$, substitute them in Eq. (\ref{eq:nu}) and calculate the reshift factor.

\bibliography{2019_ROUPAS_GP_MaxEnt}
\bibliographystyle{unsrt}

\end{document}